\documentclass[aps,prl,reprint,groupedaddress,showpacs]{revtex4-1}

\usepackage{graphicx}
\usepackage{dcolumn}
\usepackage{bm}

\begin{document}


\title{Non-equilibrium Fluctuation Relations in a Quantum Coherent
Conductor}

\author{Shuji Nakamura$^{1\dagger}$, Yoshiaki Yamauchi$^{1\dagger}$,
Masayuki Hashisaka$^1$, Kensaku Chida$^1$, Kensuke Kobayashi$^1$}
\email{kensuke@scl.kyoto-u.ac.jp} 
\author{Teruo Ono$^1$ Renaud Leturcq$^2$, Klaus Ensslin$^3$, Keiji Saito$^4$,
Yasuhiro Utsumi$^5$, Arthur C. Gossard$^6$}

\affiliation{$^1$Institute for Chemical Research, Kyoto University, Uji,
Kyoto 611-0011, Japan.}

\affiliation{$^2$Institute of Electronics, Microelectronics and
Nanotechnology, CNRS-UMR 8520, Department ISEN, Avenue Poincar{\'e}
59652 Villeneuve d'Ascq, France.}

\affiliation{$^3$Solid State
Physics Laboratory, ETH Z\"{u}rich, CH-8093 Z\"{u}rich, Switzerland.}

\affiliation{$^4$Graduate School of Science, University of Tokyo, Tokyo
113-0033, Japan.}

\affiliation{$^5$Institute for Solid State Physics, University of Tokyo,
Kashiwa, Chiba 277-8581, Japan.}

\affiliation{$^6$Materials Department, University of California, Santa
Barbara, California 93106, USA,}

\affiliation{$^{\dagger}$these authors equally contributed to this work.}

\begin{abstract}
We experimentally demonstrate the validity of non-equilibrium
fluctuation relations by using a quantum coherent conductor.  In
equilibrium the fluctuation-dissipation relation leads to the
correlation between current and current noise at the conductor, namely,
Johnson-Nyquist relation.  When the conductor is voltage-biased so that
the non-linear regime is entered, the fluctuation theorem has predicted
similar non-equilibrium fluctuation relations, which hold true even when
the Onsager-Casmir relations are broken in magnetic fields. Our
experiments qualitatively validate the predictions as the first evidence
of this theorem in the non-equilibrium quantum regime.
\end{abstract}

\date{\today}



\maketitle The fluctuation-dissipation relation~\cite{CallenPR1951} is a
central concept in physics. It was triggered by the description of
Brownian motion by Einstein~\cite{EinsteinAP1905}, who claimed that the
response of a physical system to an external force is proportional to
its equilibrium fluctuation. In electrical circuits this fact manifests
itself as the Johnson-Nyquist (JN)
relation~\cite{JohnsonPR1928,NyquistPR1928} in such a way that the
conduction through a conductor is proportional to its current noise in
equilibrium. The linear response
theory~\cite{GreenJCP1952_2,KuboJPSJ1957} founded on this relation
together with the Onsager-Casimir reciprocal
relations~\cite{OnsagerPR1931_2} provides a powerful tool to describe a
variety of physical systems. This is, however, justified only when the
systems are close to equilibrium, and the generalization to the
non-equilibrium regime has been long sought.

Generally the current ($I$) passing through a conductor can be expressed
as a polynomial of the bias voltage ($V$) as follows;
\begin{equation}
 I = G_1V + \frac{1}{2!} G_2 V^2 + \frac{1}{3!} G_3 V^3 + \cdots,
\end{equation}
where the first term represents Ohm's law with conductance $G_1$ (Such
an expansion in $V$ is assumed to be valid for mesoscopic
transport~\cite{BlanterPR2000}).  While in a mesoscopic conductor $G_1$
is directly related to the transmission~\cite{LandauerIBM1957}, the
higher-order coefficients ($G_2$, $G_3$,...) convey information on
electron-electron interactions in a voltage-biased conductor as shown
recently~\cite{SanchezPRL2004,SpivakPRL2004,WeiPRL2005, LeturcqPRL2006}.
Similarly to Eq.~(1), a polynomial of $V$ for the current noise power
$S$ generated in the conductor, namely the variance of $I$, can be
expressed as;
\begin{equation}
S = S_0 + S_1V + \frac{1}{2!} S_2 V^2 + \cdots,
\end{equation}
The coefficients of the first terms in Eqs.~(1) and (2) are linked by
the JN relation $S_0 = 4 k_BTG_1$, where $k_B$ is the Boltzmann constant
and $T$ is the temperature of the
conductor~\cite{JohnsonPR1928,NyquistPR1928,NoteOnEq2}. Now, an
essential question arises: Are there also correlations between the
higher orders (in voltage) in current $I$ and noise power $S$?

Here we demonstrate a non-equilibrium fluctuation relation in a quantum
coherent regime. We find a proportionality between $S_1$ and $G_2$,
which corresponds to the next order correlation beyond the JN relation
between $S_0$ and $G_1$.  The present relation is asymmetric in magnetic
fields unlike the Onsager-Casimir symmetry and is valid even in the
presence of interactions.  Such a direct link between non-linearity in
voltage and fluctuations out-of-equilibrium has been predicted
theoretically~\cite{TobiskaPRB2005,SaitoPRB2008,ForsterPRL2008,Forster_arXiv,AndrieuxNJP2009}
by applying the fluctuation theorem~\cite{EvansPRL1993} to a quantum
coherent conductor.  Our experiments qualitatively validate the
predictions but disagree on the quantitative level.

We used an Aharonov-Bohm (AB) ring as a typical coherent conductor.
Figure~1(a) shows an atomic force microscope (AFM) image of the AB ring
fabricated by local oxidation using an AFM~\cite{HeldAPL1998} on a
GaAs/AlGaAs heterostructure two-dimensional electron gas (2DEG) (the
electron density $3.7 \times 10^{11}$ cm$^{-2}$, the mobility $2.7
\times 10^5$ cm$^2$/Vs, and the electron mean free path $2.7$~$\mu$m at
zero back-gate voltage) as well as the experimental setup for the
two-terminal measurement in a dilution refrigerator. As our 2DEG has a
back gate to tune the electron density, the conductance of the AB ring
can be modulated by the back gate voltage $V_g$ and the magnetic field
$B$ (the AB effect). Figure~1(b) shows the conductance as a function of
$V_g$ and $B$, displaying clear AB oscillations with an oscillation
period being 25~mT, in agreement with the ring radius of
230~nm~\cite{LeturcqPRL2006}. The conductance of the ring at $B=0$~T and
$V_g =0$~V is 3.1 in units of $e^2/h \sim (25.8$~k$\Omega)^{-1}$.  The
visibility of the AB oscillations around $B=0$~T is 0.13 (see
Fig.~3(a)).

In addition to the DC measurement, we performed a noise measurement as
follows (see Fig.~1(a)). The voltage fluctuation across the sample on
the resonant circuit, whose resonance frequency is about 3.0~MHz with
the bandwidth of $\sim 140$~kHz, is extracted as an output signal of the
cryogenic
amplifier~\cite{de-PicciottoNature1997,DiCarloRSI2006,HashisakaPRB2008}. The
time-domain signal is then captured by a two-channel digitizer, and is
converted to spectral density data via FFT, where the cross-correlation
technique is adopted to increase the resolution.  By numerically fitting
the obtained resonance peak, the current noise power spectral density
$S$ is deduced as performed in Ref.~\cite{DiCarloRSI2006}.

\begin{figure}[tbp]
\center
\includegraphics[width=.93\linewidth]{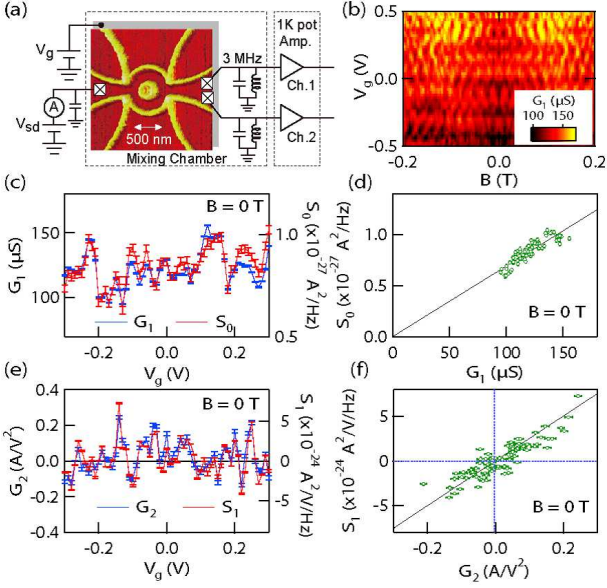}
\caption{ (a) AFM image of the AB ring with the DC and noise measurement
setup in the dilution refrigerator whose base temperature is 45 mK.  The
in-plane gates defined by the oxide lines are grounded in this
experiment.  (b) Conductance of the AB ring as a function of $V_g$ and
$B$.  (c) $G_1$ (left axis) and $S_0$ (right axis) as a function of
$V_g$ at $B=0$~T.  (d) $S_0$ is plotted as a function of $G_1$. The
solid line indicates the JN relation of $S_0 = 4 k_BTG_1$ with $T=
125$~mK. (e) $G_2$ (left axis) and $S_1$ (right axis) as a function of
$V_g$.  (f) $S_1$ is plotted as a function of $G_2$. The solid line is
the result of the fitting ($S_1 = 3.64 \times 4k_BTG_2$ with $T=
125$~mK).}
\end{figure}

We deduce the coefficients in Eqs.~(1) and (2) by numerically fitting
the measured current $I$ and current noise power spectral density $S$ as
polynomials of $V$~\cite{comment_fitting}.  In the analysis we set the
bias window to $|eV| \leq 50$~$\mu$eV, where Joule heating is negligible
as confirmed in previous noise measurements~\cite{HashisakaPRB2008} (Two
examples of the analysis are shown in Fig.~2). Figure~1(c) shows the
conductance $G_1$ (left axis) and the equilibrium noise power density
$S_0$ (right axis) as a function of $V_g$ at $B=0$~T.  Due to electron
interferences, $G_1$ varies as $V_g$ changes. The behavior of $S_0$
perfectly follows that of $G_1$ as expected from the JN relation. The
proportionality between $G_1$ and $S_0$ shown in Fig.~1(d) indicates
that $S_0 = 4 k_BTG_1$ is fulfilled with an electron temperature of $T=
125$~mK.

Now let us focus on the next order in voltage $V$. Figure~1(e) shows
$G_2$ (left axis) and $S_1$ (right axis) as a function of $V_g$.  Both
$G_2$ and $S_1$ vary reflecting electron interferences.  Moreover, their
variations are correlated (Fig.~1(f)) with a correlation factor being
0.88 between the two. When the result is expressed in the form $S_1 =
4k_BTG_2\alpha^0$ (``0'' denotes the zero-field) based on the assumption
that the functional form is similar to the JN relation, we obtain
$\alpha^0 = 3.64 \pm 0.54$ by a numerical
analysis~\cite{PassingJCCCB1983}, where the error bar indicates the 95\%
confidence interval (see the solid line in Fig.~1(f)). While the
magnitude of $\alpha^0$ will be discussed later, two remarks are made.
First, a proportionality between $G_2$ and $S_1$ is already contained in
the scattering theory for non-interacting systems~\cite{BlanterPR2000},
which yields general results to express the current and its noise as a
function of the bias voltage, the transmission of the conductor, and the
Fermi distribution of the reservoirs at a given temperature. However, we
will show later that the present relation is valid beyond the
description based on this picture in terms of the breaking of the
Onsager-Casimir symmetry.  Second, the finite values of $S_1$ indicate
that there is a certain bias voltage range where we measure ``negative
excess noise'', namely the non-equilibrium noise smaller than the
equilibrium
one~\cite{Forster_arXiv,LesovikZPB1993,BlanterPR2000}. Indeed, depending
on the coefficients ($S_1$, $S_2$,...) in Eq.~(2), the value of the bias
voltage to give the minimum noise power is shifted from zero by a value
of the order of 10 $\mu$V~\cite{Forster_arXiv} as shown in Fig.~2.

\begin{figure}[tbp]
\center \includegraphics[width=.95\linewidth]{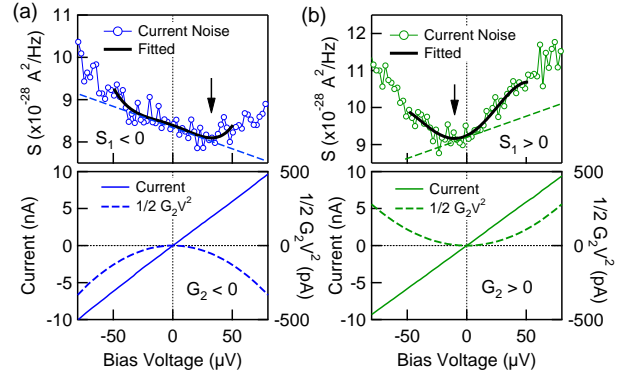}
\caption{(a) (b) (Lower panels) Two examples of $I$-$V$ characteristics
obtained at different $V_g$'s, where $\frac{1}{2}G_2V^2$ components
deduced from the numerical fitting are superposed as dashed curves.
(Upper panels) The corresponding $S$, where the results of the
polynomial fitting and $S_1V$ components are superposed as solid and
dashed curves, respectively.  The non-equilibrium noises are smaller
than the equilibrium ones at certain bias voltages indicated by the
arrows. The signs of $G_2$ and $S_1$ are correlated (negative in (a) and
positive in (b)). }
\end{figure}

The Onsager-Casimir reciprocity is a fundamental relation to
characterize the response of a system close to
equilibrium~\cite{OnsagerPR1931_2}. Figure~3(a) shows $G_1$ (left axis)
and $S_0$ (right axis) as a function of $B$ at $V_g = 0.02$ V. Both of
them, displaying a periodic oscillations due to the AB effect, are
symmetric with regard to $B = 0$~T, namely $G_1(B)=G_1(-B)$ and
$S_0(B)=S_0(-B)$. Furthermore, they are proportional to each other
(Fig.~3(b)). This is the direct consequence of the JN relation and the
Onsager-Casimir reciprocity. On the other hand, the next order
coefficient $G_2$ has no such symmetry as shown in Figs.~3(c) and 3(e)
(blue curves), where the symmetric components $G_2^S (B) \equiv G_2 (B)
+ G_2 (-B)$ and anti-symmetric components $G_2^A (B) \equiv G_2 (B) -
G_2 (-B)$ are presented, respectively.  The emergence of a finite
$G_2^A$ at $B \neq 0$ is in agreement with previous works on the
breaking of the Onsager-Casimir reciprocity in the non-linear transport
regime~\cite{SanchezPRL2004,SpivakPRL2004,WeiPRL2005,LeturcqPRL2006}. It
was attributed to electron-electron interactions in mesoscopic
conductors~\cite{SanchezPRL2004,SpivakPRL2004} since, in the Landauer
picture for non-interacting electrons~\cite{BlanterPR2000}, the
transmission probability of electrons always obeys the Onsager-Casimir
reciprocity regardless of applied bias voltages, leading to a vanishing
anti-symmetric term $G_2^A$.

\begin{figure}[tbp]
\center \includegraphics[width=.95\linewidth]{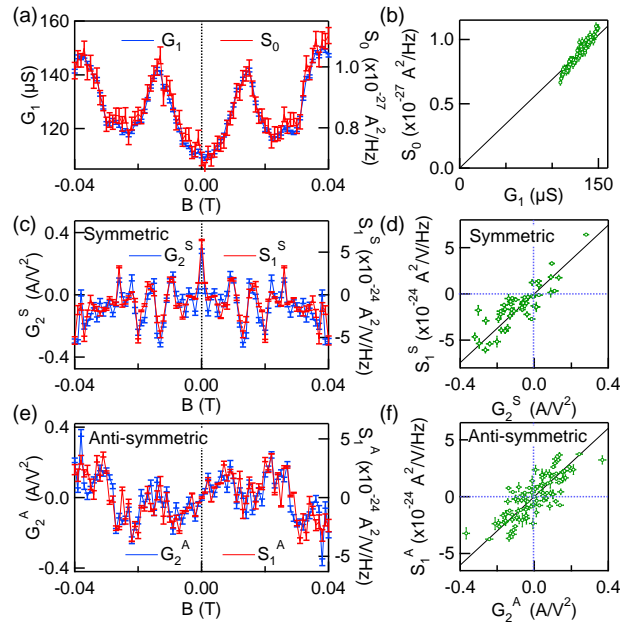}
\caption{ (a) $G_1$ (left axis) and $S_0$ (right axis) as a function of
$B$ at $V_g = 0.02$ V.  (b) $S_0$ as a function of $G_1$. The solid line
$S_0 = 4 k_BTG_1$ is the JN relation at $T= 125$~mK.  (c) Symmetric
components in the second-order coefficients $G_2^S$ (left axis) and
$S_1^S$ (right axis) as a function of $B$.  (d) $S_1^S$ as a function of
$G_2^S$. The solid line is the result of the fitting corresponding to
$S_1^S = 2.69\times 4 k_BTG_2^S$ for $T= 125$~mK.  (e) Anti-symmetric
components in the second-order coefficients $G_2^A$ (left axis) and
$S_1^A$ (right axis) as a function of $B$.  (f) $S_1^A$ as a function of
$G_2^A$. The solid line is the result of the fitting corresponding to
$S_1^A = 2.18\times 4k_BTG_2^A$ for $T= 125$~mK.}
\end{figure}

Here we show that the non-equilibrium relation between $G_2$ and $S_1$
is valid even when the Onsager-Casimir relation is broken.  In
Figs.~3(c) and 3(e) the symmetric components $S_1^S (B)\equiv S_1 (B) +
S_1 (-B)$ and anti-symmetric components $S_1^A (B)\equiv S_1 (B) - S_1
(-B)$ as a function of $B$ are superposed in red curves,
respectively. In addition to the proportionality between $G_2^S$ and
$S_1^S$ (see Fig.~3(d)), there exists a clear proportionality between
$G_2^A$ and $S_1^A$ as shown in Fig.~3(f). The correlation factors are
0.94 and 0.88 for the symmetric and anti-symmetric parts,
respectively. For the symmetric part, $\alpha^S = 2.69^{+0.59}_{-0.35}$
is obtained in the expression of $S_1^S = 4k_BTG_2^S\alpha^S$, while
$\alpha^A = 2.18^{+0.32}_{-0.18}$ is deduced in the anti-symmetric
relation $S_1^A = 4k_BTG_2^A\alpha^A$.  Importantly, the anti-symmetric
part is non-trivial in that it is the consequence of the departure of
the system from the Onsager-Casimir symmetry, and a signature of
non-equilibrium and non-linearity. Nevertheless, the higher-order
correlation still exists there as well as in the symmetric part, but
with different coefficients.

What is the reason for the correlations in the non-linear transport
regime? Here we give an intuitive picture of higher order correlations
following recent
theories~\cite{TobiskaPRB2005,SaitoPRB2008,ForsterPRL2008,Forster_arXiv,AndrieuxNJP2009}. Electron
transport can be viewed as the electron exchange process between two
reservoirs via a conductor. Consider the probability $P(Q)$ that one
reservoir gains $Q$ electrons from the other as a consequence of the
exchange (in the absence of a magnetic field, for simplicity). As time
reversal symmetry, particle-number conservation, and energy conservation
are required over the whole system including the reservoirs, $P(Q)$
should satisfy $P(Q) = P(-Q)
\exp(\frac{eV}{k_BT}Q)$~\cite{TobiskaPRB2005,SaitoPRB2008,ForsterPRL2008,Forster_arXiv,AndrieuxNJP2009}.
This means that the probability of a $Q$-electrons exchange process and
that of its time-reversed process ($-Q$-electrons exchange) are linked
with a factor difference of $\exp(\frac{eV}{k_BT}Q)$, reflecting a
probability difference of the initial states for the two exchange
processes. Note that this equality based on the micro-reversibility is
valid even in the non-equilibrium situation and provides the basis of
the fluctuation theorem~\cite{EvansPRL1993}. As $P(Q)$ contains all the
information associated with electron transport through the conductor,
the strong constraint posed by this equality on the cumulants of $Q$
yields a number of correlations starting with the JN relation. For the
second order, the equality predicts $S_1 =
2k_BTG_2$~\cite{TobiskaPRB2005,SaitoPRB2008,ForsterPRL2008,Forster_arXiv,AndrieuxNJP2009}.
In the presence of magnetic fields, the fluctuation relations that do
not rely on the microreversibility out of equilibrium were recently
pointed out~\cite{ForsterPRL2008,Forster_arXiv}, where the
anti-symmetric relation is expressed by $S_1^A-2k_BTG_2^A =
C_0^A/3k_BT$. Here, $C_0^A$ is the anti-symmetric part of the third
cumulant (``skewness'') at equilibrium. With the restriction of
micro-reversibility, this anti-symmetric relation~\cite{SaitoPRB2008} is
reduced to $S_1^A = 6k_BTG_2^A$ with another relation $S_1^A
=C_0^A/2k_BT$. In both cases, the symmetric relation is given by $S_1^S
= 2k_BTG_2^S$.

\begin{figure}[tbp]
\center
\includegraphics[width=.99\linewidth]{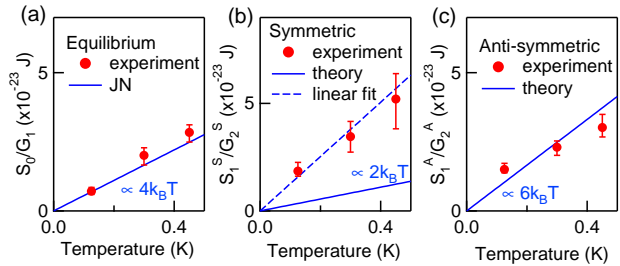}
\caption{ (a) Values of $S_0/G_1$ obtained at 125, 300, and 450 mK. The
solid line shows the JN relation.  (b) Values of $S_1^S/G_2^S$ at 125,
300, and 450 mK. The solid line is the theoretical expectation
($S_1^S/G_2^S = 2k_B
T$)~\cite{TobiskaPRB2005,SaitoPRB2008,ForsterPRL2008,Forster_arXiv,AndrieuxNJP2009}.
The dashed line is the result of the linear fitting that gives
$S_1^S/G_2^S = 4.6 \times 2 k_B T$. (c) Values of $S_1^A/G_2^A$ at 125,
300, and 450 mK. The solid line is the theoretical expectation
($S_1^A/G_2^A = 6k_B T$)~\cite{SaitoPRB2008,AndrieuxNJP2009}. }
\end{figure}

At zero-field, the observed linearity between $S_1$ and $G_2$
qualitatively agrees with the above discussion, although the theory does
not well reproduce the obtained ratio $S_1/2k_BTG_2 = 2\alpha^0 \sim 7$.
At finite magnetic field as well, while the expected proportionality is
obtained, the ratio ($S_1^S/2k_BTG_2^S = 2\alpha^S \sim 5$) is larger
than in theory. The present observation for $\alpha^0$ and $\alpha^S$ is
perfectly reproducible in our experiments performed in several different
conditions for different magnetic fields and gate voltages.  Concerning
the temperature dependence, although the ratio $S_0^S/G_1^S$ follows the
JN relation (Fig.~4(a)), the ratio $S_1^S/G_2^S$ shown in Fig.~4(b)
depends linearly on temperature as expected, but the slope is similarly
larger than the theoretical prediction. This implies the necessity of
further theoretical work that takes realistic situations into
account~\cite{UtsumiCM}. In contrast, the result for the anti-symmetric
component ($S_1^A/6k_BTG_2^A \sim 1.5$) is in better agreement with
theory $S_1^A = 6k_BTG_2^A$ in spite of a possible independent
contribution of $C_0^A$ when only the universal relation without
micro-reversibility ($S_1^A-2k_BTG_2^A = C_0^A/3k_BT$) is
assumed~\cite{ForsterPRL2008,Forster_arXiv}. The agreement is also clear
in the temperature dependence of the ratio $S_1^A/G_2^A$ as shown in
Fig.~4(c).  This observation suggests that micro-reversibility in
magnetic field is likely to be validated in the present experiment.

In conclusion we have experimentally proven the presence of the direct
link between the non-linear response and the non-equilibrium fluctuation
in the AB ring, as theoretically predicted on the basis of the
fluctuation theorem. While our demonstration was performed for the
simplest case in a normal coherent conductor without energy relaxation
in the scattering region, the present finding will further open an
applicability of the fluctuation theorem in the quantum coherent regime
and in the presence of magnetic fields.

We appreciate fruitful discussions from Markus B{\"u}ttiker, Masahito
Ueda and Takeo Kato. This work is partially supported by KAKENHI, Yamada
Science Foundation, SCAT, Matsuo Science Foundation, Strategic
International Cooperative Program the Japan Science and Technology
Agency (JST), and the German Science Foundation (DFG).

\appendix*
\section{Appendix: Simple deduction of the higher order correlations}
We explain the fluctuation theorem using the simplest setup.  We
consider a mesoscopic conductor, say a quantum point contact, where the
two quantum wires are coupled by tunneling. For simplicity, no magnetic
field is applied.  The present system is described by the following
Hamiltonian
\begin{eqnarray}
H &=& H_L + H_R + H_{LR},
\end{eqnarray}
where $H_L$ and $H_R$ are the Hamiltonian of the left and right quantum
wires and $H_{LR}$ is the tunneling part between them.  The initial
density matrix is decoupled into the equilibrium states of each wire,
where the left and right wires are assumed to have equal temperature
$1/\beta$ and have chemical potentials $\mu_L$ and $\mu_R$,
respectively. Then the whole density matrix is described by
\begin{eqnarray}
\hat{\rho}_{\rm initial} &=& \sum_{n_L , n_R }  \rho_{n_L, n_R }
|n_L , n_R\rangle \langle n_L , n_R |  , \\ 
\rho_{n_L, n_R } &=& 
{e^{-\beta [E_{n_L} - \mu_{L} \, n_{L } ] } \over Z_{L}}
{e^{-\beta [E_{n_R} - \mu_{R} \, n_{R } ] } \over Z_{R}},
\end{eqnarray}
where $Z_L$ and $Z_R$ are the normalization factors, and $|n_L , n_R
\rangle$ defines the state that $n_L$ and $n_R$ electrons are present
inside the left and right wires with the eigenenergies $E_{n_L}$ and
$E_{n_R}$ of $H_L$ and $H_R$, respectively.  The probability to find the
state $|n_L ' , n_R '\rangle $ after a certain time $\tau$ starting from
the initial state $|n_L , n_R \rangle$ is expressed as
\begin{eqnarray*}
P_{(n_L , n_R )\to (n_L' , n_R')} 
\!=\! |\langle n_L' , n_R' | e^{{-i\tau \over \hbar}H } | n_L , n_R \rangle |^2 
\rho_{n_L , n_R} . ~
\end{eqnarray*}
We note the time reversal symmetry
\begin{eqnarray*}
|\langle n_L' , n_R' | e^{{-i\tau \over \hbar } H } | n_L , n_R \rangle |^2  \!=\!
|\langle n_L , n_R | e^{{-i\tau \over \hbar }  H } | n_L ', n_R '\rangle |^2  .~
\end{eqnarray*}
We also note electron number conservation $n_{L}- n_L' = -(n_R - n_R') $
and energy conservation satisfied at very large $\tau$: $E_{n_L'}
-E_{n_L} \approx -(E_{n_{R}' } - E_{n_R } ) $. Using these time
reversal symmetry and conservation laws, we find the relation
\begin{eqnarray*}
P_{(n_L ,n_R)\to (n_L' , n_R') } 
&=& P_{(n_L' ,n_R')\to (n_L , n_R) }  e^{A (n_L - n_{L}')}, \label{dft}
\end{eqnarray*}
where $A$ is an affinity $A=\beta (\mu_L - \mu_R)$.  The probability
that transmitted electron number is $Q$, is defined as $P(Q) = \sum_{n_L
, n_R , n_L' , n_R'}P_{(n_L ,n_R)\to (n_L' , n_R') } \delta (Q - (n_L -
n_{L}'))$.  A direct consequence from the relation (\ref{dft}) is
``Fluctuation Theorem'':
\begin{eqnarray}
P(Q) &=& P(-Q) e^{AQ} \label{ft} .
\end{eqnarray} 

Now let us discuss the higher order correlations between the current and
its noise power, which are addressed in the present paper.  With
Fluctuation Theorem (\ref{ft}), we find the following identity
\begin{eqnarray}
\langle Q \rangle &=& \int d Q Q P(Q) =
- \int d Q Q P(Q) e^{-A Q} \nonumber \\
&=&
-\langle Q \rangle + A \langle Q^2 \rangle -{A^2\over 2! }\langle Q^3 \rangle +
\cdots . 
\end{eqnarray}
Furthermore, we note that $\langle Q^n \rangle$ is a function of $A$, i.e.,
\begin{eqnarray}
\langle Q^n \rangle &=& \langle Q^n \rangle_0 + A \langle Q^n \rangle_1
+ {A^2\over 2!} \langle Q^n \rangle_2 + \cdots .
\end{eqnarray}
Comparing order by order with respect to $A$, we find infinite number of 
relationships among these quantities, some of which are given as
\begin{eqnarray}
\langle Q^2 \rangle_0 &=& 2 \langle Q \rangle_1 , \label{1st}\\
 \langle Q^2 \rangle_1  &=& \langle Q \rangle_2  . \label{2nd}
\end{eqnarray}
Average current $I$ and current noise power $S$ are defined as $I=\langle Q
\rangle /\tau$ and $S = 2(\langle Q^2 \rangle - \langle Q \rangle^2
)/\tau$.  The first relation (\ref{1st}) is equivalent to the
fluctuation dissipation $S_0 = 4 k_{\rm B} T G_1$ [G. Gallavotti, {\it
Phys. Rev. Lett.} {\bf 77,} 4334 - 4337 (1996)], and the second relation
(\ref{2nd}) is to $S_1 = 2 k_{\rm B} T G_2$ (See main text for the
definitions of $G_1$, $G_2$, $S_0$, and $S_1$).  More systematic and
exact derivation for these relations including the finite magnetic field
case is performed by using a cumulant generating function [K. Saito \&
Y. Utsumi, {\it Phys. Rev. B} {\bf 78,} 115429 (2008)].


\begin{thebibliography}{99}
\bibitem{comment_fitting} The polynomial fitting was performed by taking
up to the fifth order of $V$ for $I$ and up to the fourth order of $V$
for $S$ into account, respectively. The analysis up to third or seventh
order of $V$ for $I$ and second order of $V$ for $S$ yields results
consistent with those presented here.
\end{thebibliography}
\end{document}